\documentclass[english,aps,prb,letterpaper,manuscript,floatfix,superscriptaddress]{revtex4}
\usepackage[T1]{fontenc}
\usepackage{color}
\usepackage[latin1]{inputenc}
\usepackage{graphicx}
\usepackage{bm}
\usepackage{epsfig}
\usepackage{rotating}
\usepackage{float}

\usepackage[normalem]{ulem}

\usepackage{dcolumn}
\usepackage{bm}

\begin{document}

\title{\bf Molecular ordering in lipid monolayers: an atomistic simulation}
\author{S. Panzuela}
\email{sergio.panzuela@uam.es}
\affiliation{Departamento de F\'{\i}sica Te\'orica de la Materia Condensada,
Universidad Aut\'onoma de Madrid,
E-28049, Madrid, Spain}

\author{D. P. Tieleman}
\email{tieleman@ucalgary.ca}
\affiliation{Centre for Molecular Simulation and Department of Biological Sciences, University of Calgary, Calgary, Alberta, Canada}

\author{L. Mederos$^*$}
\email{lmederos@icmm.csic.es}
\affiliation{Instituto de Ciencia de Materiales de Madrid, Consejo Superior de Investigaciones
Cient\'{\i}ficas, C/Sor Juana In\'es de la Cruz, 3, E-28049, Madrid, Spain}

\author{E. Velasco}
\email{enrique.velasco@uam.es}
\affiliation{Departamento de F\'{\i}sica Te\'orica de la Materia Condensada,
Instituto de Ciencias de Materiales Nicol\'as Cabrera, and IFIMAC,
Universidad Aut\'onoma de Madrid, E-28049, Madrid, Spain}

\date{\today}

\begin{abstract}  
We report on atomistic simulations of DPPC lipid monolayers using the CHARMM36 lipid force field and four-point OPC water model. The entire two-phase region where
domains of the `liquid-condensed' (LC) phase coexist with domains of the `liquid-expanded' (LE) phase has been explored. The simulations are long enough that 
the complete phase-transition stage, with two domains coexisting in the monolayer,
is reached in all cases. Also, system sizes used are larger than in previous works.
As expected, domains of the minority phase are elongated, emphasizing
the importance of anisotropic van der Waals and/or electrostatic dipolar interactions
in the monolayer plane. The molecular structure is quantified in terms of distribution functions for the hydrocarbon chains and the PN dipoles. In contrast to previous work, where average distributions are calculated, distributions are here extracted for each of the coexisting phases by first identifying lipid molecules that belong to either LC or LE regions. The three-dimensional distributions show that the average tilt angle of the chains with respect to the normal outward direction is $(39.0\pm 0.1)^{\circ}$ in the LC phase.
In the case of the PN dipoles the distributions indicate a tilt angle of $(110.8\pm 0.5)^{\circ}$ in the LC phase and $(112.5\pm 0.5)^{\circ}$ in the LE phase. These results are quantitatively different from previous works, which indicated a smaller normal component of the PN dipole. Also, the distributions of the monolayer-projected chains and PN dipoles have been calculated. Chain distributions peak along a particular direction in the LC domains, while they are uniform in the LE phase. Long-range ordering associated with the projected PN dipoles is absent in both phases. 
These results strongly suggest that LC domains do not exhibit dipolar ordering in the plane of the monolayer, the effect of these components being averaged out
at short distances. Therefore, the only relevant component of the molecular 
dipoles, as regards both intra- and long-range interdomain interactions, is normal to the monolayer. Also, the local orientation
of chain projections is almost constant in LC domains and points in
the direction along which domains are elongated,
suggesting that the line tension driving the phase transition is anisotropic with respect to the interfacial domain boundary. Both van der Waals interactions
and interactions from normal dipolar components 
seem to contribute to an anisotropic line tension, with dipolar in-plane 
components playing a negligible role. 
\end{abstract}

\maketitle

\section{Introduction}

Langmuir monolayers of DPPC in the air-water interface have long been considered a fruitful model system to understand basic physical properties of lung
surfactant\cite{Perejil1,Perejil2,Casals}. A huge body of experimental work has been accumulated over the last decades on this system, focusing on the thermodynamics (surface pressure 
isotherms), molecular structure, and domain shape and structure \cite{Kaganer,Peter1,Peter2}.
One of the puzzling questions about DPPC monolayers 
is the possibly nonequilibrium phenomena involved in the growth of domains in the two-phase coexistence region between the liquid-condensed (LC) and liquid-expanded (LE) phases. The difficulty to
observe a truly horizontal sector in surface-pressure--area isotherms \cite{isotherm}, together with the observation of persistent, seemingly equilibrated domains with
uniform size throughout the coexistence region, may indicate the presence of very long relaxation processes and long-range domain interactions \cite{domains1,domains2,domains3,domains4}. 
Also, some theoretical models have been 
developed to explain domain shapes and transitions between different shape r\'egimes. These models contain effects from different types of interactions:
line tension, curvature, and dipolar components in directions normal and perpendicular to the monolayer. Competition between these interactions
leads to interesting shape behaviour, in some cases close to experimental observations of domains in monolayers of pure DPPC molecules.
Detailed experimental work \cite{Mohwald,Ma} and computer simulations \cite{Dominguez,Rose,Bresme,Huynh,Javanainen} also exist which suggest
models for molecular arrangement of lipid molecules in the monolayers, a question which is also open to debate.
The very nature of the relevant interactions is uncertain. 
A broad consensus exists on the PN dipole orientation, which most studies interpret to be more or less
parallel to the monolayer \cite{Hauser2,Dominguez,Gallegos}. Also, it has been postulated that the PN dipole is located
at slightly different depths in the liquid subphase: as the film is compressed, 
molecules condense into LC regions and the entire lipid molecules are displaced towards the air phase,
thereby expelling solvating water molecules and making the PN region more dehydrated \cite{Gallegos}. 
These analyses suggest a model where LC domains made of
molecules with tightly-packed, tilted (with respect to the monolayer normal) 
aliphatic chains and strongly interacting dipoles parallel to the
monolayer coexist with LE regions with disordered chains and screened parallel dipoles. This view
somehow contradicts the assumptions made by many theoretical models for domain shape and shape transitions, 
based on distributions of
strongly interacting dipoles {\it normal} to the monolayer, with van der Waals (vdW) interactions 
effectively showing up in a line-tension contribution which is {\it isotropic} with respect to 
the in-plane normal direction at the domain line boundary. The competition between an isotropic
line tension and a long-range electrostatic contribution coming from dipoles along the monolayer
normal leads to interesting behaviours. In-plane dipolar components have also been 
contemplated and seen to renormalise the line tension by making it anisotropic\cite{Sriram}. However, the question
is whether assumptions behind these theoretical models are completely correct.

The present work is an attempt to shed some light on some of these issues from the point of view of atomistic simulation, using state-of-the-art force field modelling for both lipid and water molecules, and constitutes a quantitative leap forward with respect
to previous simulations in terms of system size and simulation time\cite{Rose,Bresme,Huynh,Comment,Reply,Bresme2,Javanainen}. 
Recently, Javanainen et al. 
\cite{Javanainen} reported on full-atom simulations of DPPC monolayers.
Lipid interactions were modelled using the CHARMM36 force field\cite{CHARMM}, whereas for water molecules the four-point OPC4 force field\cite{OPC4} was used.  
It was shown that the combination of these two force fields provides quantitative agreement with experimental results on surface pressure isothems of DPPC monolayers,
making this model a suitable tool to also explore structural molecular behaviour. 
Here we use a simulation setup similar to that of
Javanainen et al. to examine in detail the structure of the monolayer
in conditions of full phase-separation, where single LC or LE domains 
coexist with the opposite phase. A novel feature of our analysis is that
various molecular distribution functions are identified separately
for the two phases, allowing for more accurate measurements of 
molecular arrangements and orientation. Also, correlation functions
describing molecular ordering in both phases are presented. These functions exhibit
features corresponding to sustained long-range order of the tails
in the LC domains, but PN dipolar heads show almost perfect disorder
in azimuthal angles, which indicates that the role of in-plane
dipolar components is negligible in determining domain shape and long-range domain
interactions.

\begin{figure}
\begin{center}
\includegraphics[width=0.80\linewidth,angle=0]{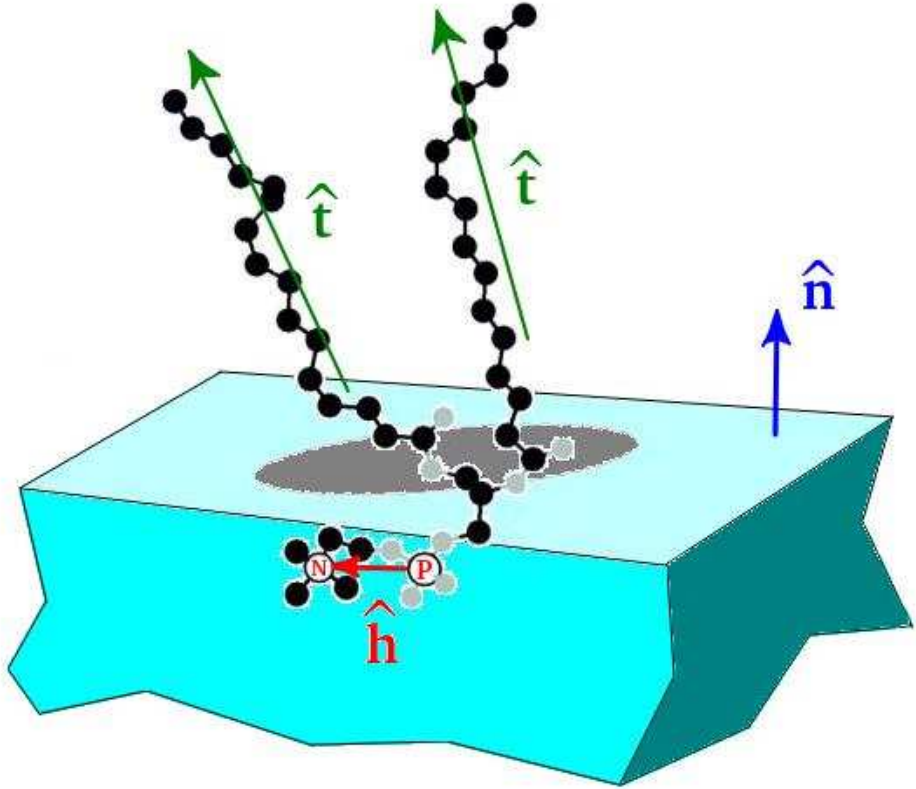}
\caption{\label{fig1} Schematic of a DPPC molecule showing the definition of the three unit vectors used to describe the 
orientation of different molecular groups. Also shown is the projection 
of the molecule on the monolayer plane (see text for details).}
\end{center}
\end{figure}

\section{Simulations}

In the present paper we use the same combination of force
fields used by Javanainen et al., but on larger system sizes.
We run NVT Molecular Dynamics (MD) simulations on systems consisting of 1152 DPPC molecules, distributed in two monolayers (576 molecules
per monolayer) and a liquid slab consisting of 93312 water molecules, creating two equivalent liquid-vapour interfaces. All systems were
built using CHARMM-GUI, a web-based graphical interface that helps create molecular input files for CHARMM-based atomistic simulations. MD simulations were
run using the GROMACS 2018.3\cite{GROMACS1,GROMACS2} software on a set of 288 CPU cores for about 21 days of CPU time. Each run spanned a total of 300 ns (simulation time), comprising 50 ns
for equilibration and 250 ns for averaging. The systems were thermostatted at 298 K using the Nos\'e-Hoover technique with coupling parameter $1.0$ ps,
and a leap-frog algorithm with timestep $0.002$ ps was used to integrate the equations of motion. 

\begin{figure}[h]
\begin{center}
\includegraphics[width=0.85\linewidth,angle=0]{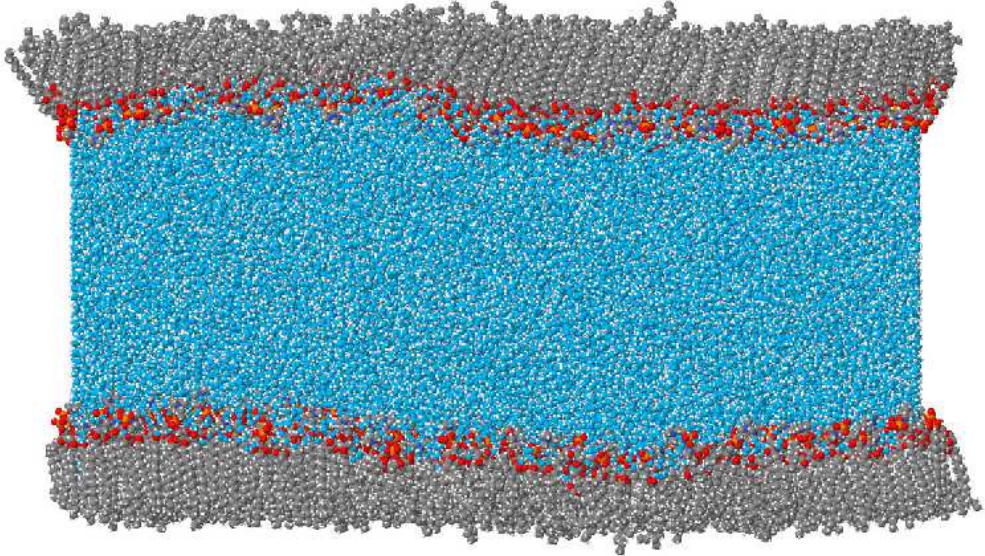}
\caption{\label{fig2} Edge-on view of
molecular configuration of a water liquid slab containing two lipid monolayers with an area per 
lipid of $65$ \AA$^{2}$. Except for small-amplitude
interfacial fluctuations, the liquid slab is parallel to the $xy$ plane. The presence of large LC
domains at both interfaces is clearly visible.}
\end{center}
\end{figure}

Fig. \ref{fig1} is a schematic of a DPPC molecule and its projection on the monolayer plane, showing the definition of the three unit vectors used to describe the orientation of different molecular groups 
($\hat{\bm h}$ for the head and $\hat{\bm t}$ for the chains; their precise definitions are given below). 
Fig. \ref{fig2} depicts a molecular configuration of the water slab and lipid
molecules adsorbed at the two interfaces, corresponding to two equivalent lipid monolayers.
During the course of the simulations no water molecules were detected in the vapour side, which is therefore a vacuum.
The monolayers are parallel to the $xy$ plane and perpendicular to the $z$ axis, except for small-amplitude
interfacial fluctuations. The simulation
box is square in the $xy$ plane with total area $A=L_xL_y$
adjusted to fit the desired two-dimensional inverse 
density, here expressed, as is customary, as area per lipid in units of \AA ngstr\"om$^{2}$ (\AA$^2$). 
The box length in the $z$ direction, $L_z$, was set to $22$ nm. The number of water molecules used 
was $93312$, giving a thickness of $15$ nm. Selected runs were conducted with $L_z=65$ nm 
in order to check for possible size effects. Also simulations of a single monolayer were performed to
reveal possible effects coming from the interaction between the monolayers mediated by the 
water slab. In no case was any significant deviation detected,
so that all results presented below were obtained with the parameters mentioned above.

Electrostatic
interactions were handled with the Particle Mesh Ewald (PME) method\cite{PME}, treating interactions explicitely up to $1.2$ nm and using Fast Fourier Transform for the long-ranged part. 
All vdW interactions were truncated and shifted at $r_c=1.2$ nm and corrections to pressure and energy were correspondingly applied.
Each run was conducted at constant lateral area and volume to explore different state points along the coexistence line of the LC-LE phase transition.
Therefore different mean areas per lipid, from $52$ to $76$ \AA$^2$, were analysed which bracket the coexistence density interval. 
We later report on improved estimates for coexistence densities
based on actual measurements of local densities inside domains.

As mentioned in the introduction, the main focus of our work is on 
domain structure at LC-LE coexistence. Therefore, a reliable method to 
distinguish lipid molecules
in LC domains from lipid molecules in LE domains is required. Since the phase
transition involves structural changes that mostly occur in the plane of the monolayer, we need to
define a two-dimensional position associated to each lipid molecule. This was done 
by calculating the centre of mass, projected on the $xy$ plane, of the C$=$O units of the 
glycerol group.
In this way we associated a two-dimensional position vector ${\bm R}=(X,Y)$ to each lipid molecule.

\begin{figure}[h]
\begin{center}
\includegraphics[width=0.85\linewidth,angle=0]{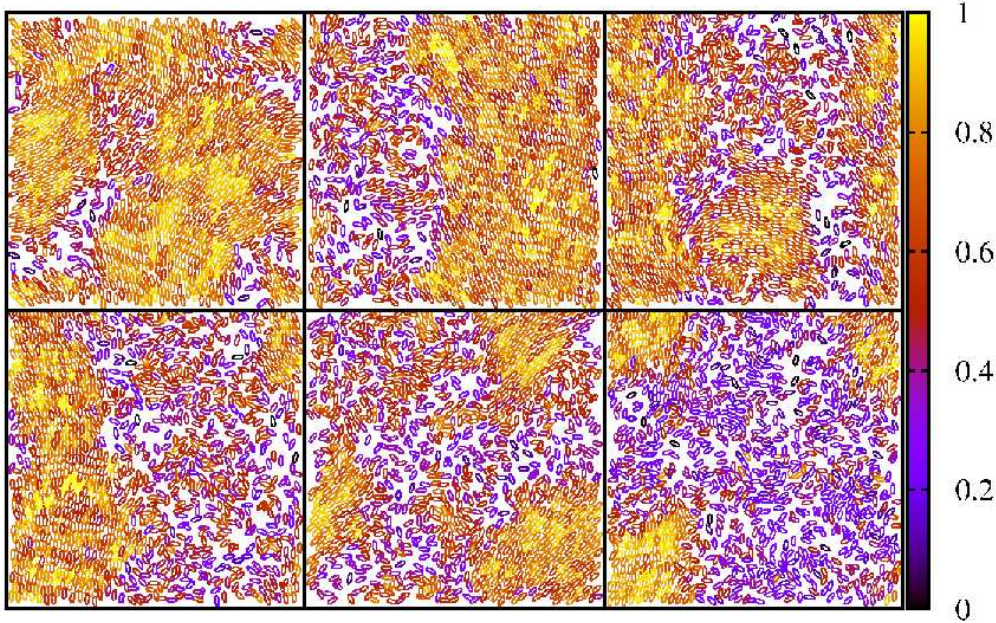}
\caption{\label{fig3} Top configurations of one of the monolayers for different values of the mean area per lipid
(from top-left panel and clockwise: $58$ \AA$^2$, $61$ \AA$^2$, $63$ \AA$^2$, 
$65$ \AA$^2$, $68$ \AA$^2$, and $71$ \AA$^2$). Lipid molecules are represented as ellipses with an 
aspect ratio of 2, oriented along the 
projection of the tail vector on the monolayer. Ellipses are coloured 
according to the value of the crystalline 
order parameter $q^{(j)}$ (scale given by the accompanying colour bar).} 
\end{center}
\end{figure}

The structural differences between LE and LC regions in the monolayer 
(see Fig. \ref{fig3}, which is explained in great detail below, for an illustration of the two regions), are well known \cite{Kaganer}.
In a LE region 
gauche defects of the two aliphatic chains of a lipid are excited, resulting in
disordered chains which interact weakly. Lipids do not
show any type of lateral order. By contrast, lipids in a LC domain exhibit
straight chains that interact strongly and present a tilted orientation
to optimise vdW interactions. As a result, lipid molecules
arrange into a stretched local triangular lattice in the plane of the monolayer
exhibiting quasi-long range order.
The stretching direction coincides with one of the 
three equivalent axes of the triangular lattice. Different methods have been implemented to distinguish these two types of
situations. A popular one is that employed by Baoukina et al. \cite{Baoukina,Peter2}, 
where a Voronoi network is 
computed and neighbour numbers are associated to each molecule.
In our case we found it more convenient to define a crystalline order parameter $q^{(j)}$ on each lipid molecule, a method borrowed 
from analyses of two-dimensional crystals. The order parameter at the $j$th molecule is defined, 
for each configuration, as
\begin{eqnarray} 
q^{(j)}=\left|\sum_{kl}e^{\displaystyle 
i{\bm K}_k\cdot({\bm R}_l-{\bm R}_j)}\right|,
\label{Qk}
\end{eqnarray} 
where $\left|\cdots\right|$ denotes the modulus of a complex number.
The sums run over all positions ${\bm R}_l=(X_l,Y_l)$ 
of the nearest neighbours of the $j$th molecule,
and the reciprocal vectors of the triangular lattice ${\bm K}_k$. 
Our simulations show that the direction along which the lattice is stretched 
coincides with the projected molecular tilt. Moreover, 
this crystallographic axis is slightly distorted inside the domain. We start 
by first obtaining the local orientation of the lattice. This is easy if we calculate the unit 
vector $\hat{\bm t}$ along each lipid chain (two per DPPC lipid) 
from the gyration tensor of the chain units (using only the carbon atoms).
Next we calculate the order tensor at the $j$th
molecule, $Q^{(j)}$, the components of which are given by
\begin{eqnarray}
Q^{(j)}_{\alpha\beta}=\frac{1}{2}\sum_{l}\left(3
\hat{t}_{\alpha}^{(l)}\hat{t}_{\beta}^{(l)}-
\delta_{\alpha\beta}\right),\hspace{0.4cm}\alpha,\beta=1,2,3
\end{eqnarray}
where the sum runs over all the neighbours located at a lateral
distance within a circle of radius 1.2 nm. The local lattice orientation is now obtained by 
calculating the eigenvector $\hat{\bm e}$ associated to the largest eigenvalue of the $Q$ tensor,
and then projecting it on the $xy$ monolayer plane. We define the corresponding normalised
vector as $\hat{\bm e}_{\perp}$.
The order parameter $q^{(j)}$ for a given lipid is in the
interval $[0,1]$, with $0$ corresponding to a totally disordered environment
and $1$ to perfect order. Even though
there is some degree of local order in the LE phase, the crystalline
order parameter $q^{(j)}$ very accurately discriminates lipids in the LC
phase from lipids in the LE phase (except perhaps close or at the 
boundary between LC and LE domains). Also, it performs better than a criterion
based on the number of neighbours (such as the Voronoi method) because $q^{(j)}$ is very sensitive to angular order of lipids
in the neighbourhood of a given lipid.

In order to describe molecular orientation with respect to the monolayer we define the following 
unit vectors: (i) $\hat{\bm n}$, normal to the monolayer, which points along the $\pm z$ axes of the 
simulation box for the top and bottom monolayers, respectively; 
i.e. $\hat{\bm n}$ is defined at each of the water liquid-vapour interfaces 
as a normal vector pointing to the vapour side.
(ii) $\hat{\bm t}$, the chain unit vector, which has been defined above.
And (iii) $\hat{\bm h}$, the PN unit vector, defined to
point from the phosphorous P atom to the nitrogen N atom of each lipid. 
We also define the projected components on the 
monolayer of these vectors as ${\bm t}_{\perp}=\hat{\bm t}-\left(\hat{\bm t}\cdot\hat{\bm n}\right)\hat{\bm n}$,
${\bm h}_{\perp}=\hat{\bm h}-\left(\hat{\bm h}\cdot\hat{\bm n}\right)\hat{\bm n}$, and their corresponding
normalised versions ${\hat{\bm t}}_{\perp}$ and $\hat{\bm h}_{\perp}$. 

With these vectors we define the distributions of the angles between the vector pairs $\{\hat{\bm t},\hat{\bm n}\}$
(distribution of the tails about the monolayer normal), $\{\hat{\bm h},\hat{\bm n}\}$ 
(distribution of the PN vector about the monolayer normal), and $\{\hat{\bm t},\hat{\bm h}\}$ 
(distribution of the angle between tail and PN vectors). All of these distributions will be called
{\it three-dimensional} (3D) distributions. In addition, we also compute the distribution of the angle
between the vectors $\{\hat{\bm t}_{\perp},\hat{\bm h}_{\perp}\}$.
This will be called {\it two-dimensional} (2D) distribution.
Both 2D and 3D distributions will be computed separately for the LC and LE regions, using the 
method described above, to understand
the different orientational properties of molecules in the two phases.

\section{Results}

\subsection{Domain identification and domain shape}

Fig. \ref{fig3} shows top views of one of the two monolayers for typical configurations at different areas per lipid,
ranging from the pure LC phase to the pure LE phase. 
Lipid chains are represented by ellipses of length-to-width 
ratio equal to 2, oriented along the projection of the tilt vector on the monolayer, and coloured according to the 
value of the crystalline order parameter at each lipid molecule. 
High values of $q^{(j)}$ are identified with the LC phase, whereas low values
correspond to LE regions. From this figure we see that, except in one case, the final separation stage of the
phase transition has been reached, since only two coexisting domains of LC and LE regions exist in the
monolayers. At high mean area per lipid LC domains are small, becoming bigger as the area per
lipid is decreased. Domains are not circularly-shaped, as would be expected in systems with 
isotropic lateral interactions, but tend to be elongated when they are large enough. In other words, domain
shapes seem to indicate that the curvature along the boundary is not constant. Also, there is a clear asymmetry 
between LC and LE domains as far as the sign of the curvature is concerned: While LC domains seem to have a preference 
for positive curvature, LE ones show negative curvatures. 
Crystallinity in LC domains appears to be quite uniform with only slight distortion of crystalline axes throughout the
domains. Also note that a simple visual inspection of Fig. \ref{fig3} seems to indicate that the projected tilt angle (i.e. the
local crystalline axes) are at an oblique angle with respect to the interface, as it is also apparent in Fig. 2 of Javanainen et 
al. \cite{Javanainen}.

\begin{figure}[h]
\begin{center}
\includegraphics[width=0.75\linewidth,angle=0]{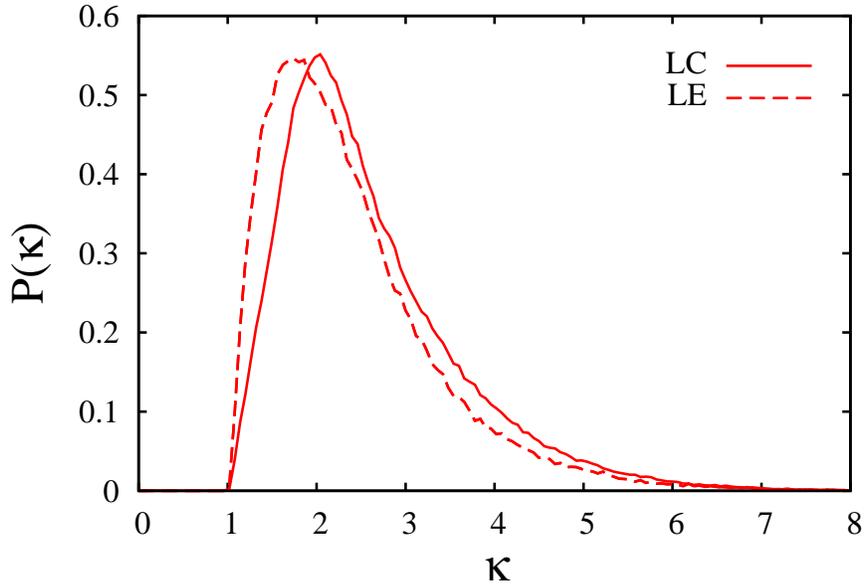}
\caption{\label{Aspect} Aspect-ratio distributions in the LC and LE domains
for a mean area per lipid of $63$ \AA$^{2}$.}
\end{center}
\end{figure}

Fig. \ref{fig3} represents the projections of the molecules as
ellipses of aspect ratio 2. The aspect ratio of a molecule on the monolayer plane is obtained by calculating
the two-dimensional gyration tensor using the projected atomic coordinates. The ratio of the two eigenvalues of this tensor
gives the aspect ratio $\kappa>1$. Fig. \ref{Aspect} shows the distribution of $\kappa$ in the LC and LE domains. As can be
seen, both distributions have a maximum around $\kappa=2$. The calculated mean aspect ratios in both regions 
are very close to 2.5.

\subsection{Coexistence densities}

\begin{figure}[h]
\begin{center}
\includegraphics[width=0.8\linewidth,angle=0]{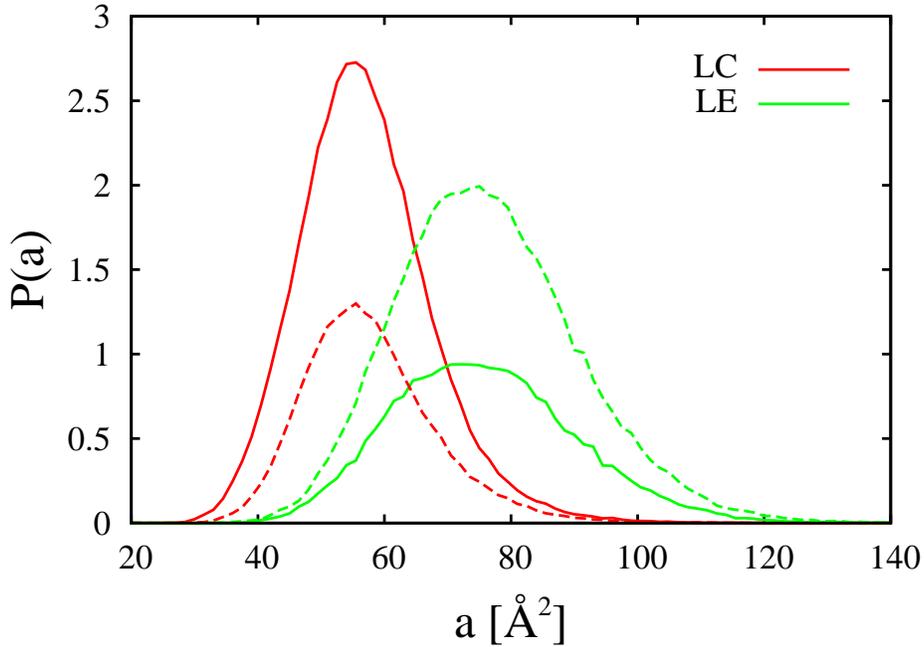}
\caption{\label{area_lipid} Histograms of local areas per lipid for the LC and LE phases.
Mean area per lipid is $61$ \AA$^2$ (continuous curves) and $68$ \AA$^2$ (dashed curves).}
\end{center}
\end{figure}

Our ability to accurately identify domains and the molecules that belong to domains allows us to calculate
the local two-dimensional density at the molecules separately in each phase. 
Local densities can be computed, regardless of the existence of domains, by counting the number of molecules
within a circle of radius 0.7 nm centered at a given molecule. 
The resulting global distribution of local densities turns out to be unimodal since local densities in 
the LE regions vary by a large amount and, therefore, coexistence densities cannot be extracted directly from such a distribution.
However, Javanainen et al. \cite{Javanainen} chose to determine the coexistence densities from a fit of the global distribution of
local density, as obtained from a Voronoi tesselation, to two Gaussian functions, each representing the distribution of densities in
one of the two coexisting phases. Instead, here we take advantage of our ability to identify domains to directly obtain the density
distribution in each phase. Fig. \ref{area_lipid} shows a histogram of the local areas per lipid for the two phases separately, 
for the cases where the
mean area per lipid is $61$ \AA$^2$ and $68$ \AA$^2$.
The two distributions overlap, with the one corresponding to the LC domains being more localised (as expected for a phase with 
a high degree of crystallinity). However, their maxima are well separated and mean values can be 
obtained separately for each histogram. These 
values are $54.8$ \AA$^2$ and $74.8$ \AA$^2$ for a mean area per lipid of $61$ \AA$^2$,
which coincide very well with the ones obtained from other values of 
mean areas per lipid. Note from the figure that the individual density distributions are not exactly Gaussians. This fact leads us to 
believe that our methodology could provide more reliable values for the coexistence densities. 

\subsection{3D distributions}

\begin{figure}[h]
\begin{center}
\includegraphics[width=1.00\linewidth,angle=0]{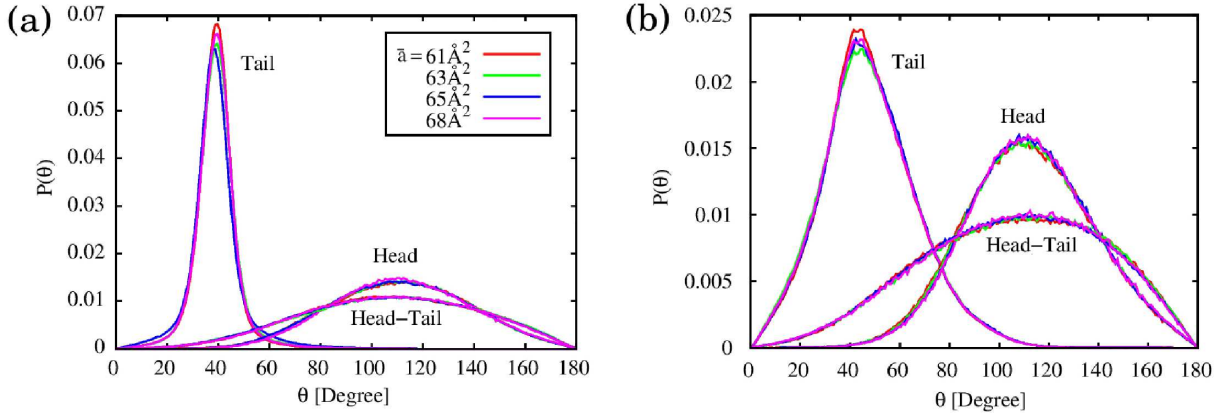}
\caption{\label{fig5} (a) 3D distributions of chain angle, PN angle, and relative chain-PN angle in LC
domains. (b) Same as in (a), but for LE domains. Mean areas per lipid $\overline{a}$
are indicated in the key.}
\end{center}
\end{figure}

The spatial arrangement of the projected tail vectors shown in Fig. \ref{fig3}
confirms that lipid chains are oriented in the LC regions.
This is of course a signature of the LC phase, which involves perfect side-to-side packing of aliphatic chains in
planar, all-gauche conformations \cite{Kaganer}. More important, visual inspection of the configurations points to global 
ordering of the tails in LC domains. Chains are not only ordered but also 
tilted with respect to the monolayer normal. The distributions in the chain angle are shown in
Figs. \ref{fig5} for LC (panel (a)) and LE (panel (b)) domains. These two distributions are largely
independent of mean area per lipid, as expected since we are probing the coexistence region
of a first-order phase transition.
The distribution is 
strongly peaked in LC regions but much broader in LE regions, consistent with the model of 
straight and packed chains in the LC molecular configurations and disordered chains in the LE phase.
From the mean of the distribution we obtain a tilt angle with respect to the outward interface normal
of $(39.0\pm 0.1)^{\circ}$ for the LC phase (standard deviation is obtained from the values for the 
different values of area per lipid).
The value for the LE phase is $(48.1\pm 0.5)^{\circ}$. This is less relevant
as the distribution is much broader because of the high orientational disorder of chains. 

The chain angle distribution has been measured previously by other authors using atomistic simulation. Choe et al. \cite{Choe} simulated
a DPPC monolayer at higher temperature (325 K) and different mean area per lipid ($56$ \AA$^2$), in the supercritical region. 
Their distribution is much more peaked than in our case, giving a mean tilt angle of $\sim 25^{\circ}$. This is not surprising, as we are
probing the tilt angle at lower temperature and density. High density monolayers are known to exhibit
a reduced tilt angle \cite{Kaganer}. The same result for the chain angle was obtained by
Dom\'inguez et. al \cite{Dominguez} for a similar temperature ($323$ K) but at a considerably
lower density (mean area per lipid of $55$ \AA$^2$). This seems to contradict the widely 
accepted idea that the tilt angle decreases as the monolayer is compressed. However, we should
bear in mind that the two works use slightly different force fields, and also that the
simulations use very small system sizes (90 and 32 lipid molecules, respectively).

PN-dipole distributions are also plotted in Figs. \ref{fig5}(a) and (b) for the LC and LE domains.
Again the dependence on mean area per lipid is weak. From the mean values we obtain tilt angles 
of $(110.8\pm 0.5)^{\circ}$ 
for the LC phase and $(112.5\pm 0.5)^{\circ}$ for the LE phase. Not only the tilt angle of the PN dipole is quite
similar in the two phases, but also the full distributions. This implies that the phase transition mainly involves a 
configurational change of the lipid chains and that the distribution of dipoles does not significantly depend on the 
local lipid density. This conclusion is consistent with the experimental studies of Hauser et al. \cite{Hauser} and of 
Ma and Allen \cite{Ma}. On the other hand, Mohammad-Aghaie et al. \cite{Bresme} report a 
bimodal PN-dipole distribution in the LC phase for a slightly lower temperature, using the
GROMOS force field and a monolayer of 64 lipid molecules. The bimodality tends to disappear as 
temperature increases. Sun \cite{Sun} reports a simulation of a monolayer with 
56 molecules of a lipid similar to DPPC (same polar head but longer tail),
using the CHARMM27 force field at a temperature of $293$ K. The PN-dipole distribution also turns 
out to be bimodal. This feature is attributed \cite{Sun} to two competing dipole orientations.
However, the origin of the two favoured orientations is not clear. Our distributions, by contrast,
show no signature of bimodality, as would correspond to a truly equilibrated LC phase
that coexist with a LE phase. Since domains of coexisting LE and LC phases are clearly 
identified in our simulation and system size is much larger 
than in previous works \cite{Bresme,Sun}, signatures of both phases are not mixed in our
simulation, and lack of relaxation to equilibrium is probably not an issue. Also note that 
our results indicate that the PN-dipole distributions in the LC and LE phases are almost identical,
as can be inferred from Figs. \ref{fig5}(a) and (b) by comparing their heights and taking into
account that both distributions are normalized.

Another very interesting distribution involves the angle between chains and head dipoles, also
shown in the figures. The results are rather surprising, since they imply very broad distributions,
even in the LC phase. Their means,
$(71.8\pm 0.6)^{\circ}$ and $(75.5\pm 0.5^{\circ})$ for LC and LE phases, respectively,
are consistent with the mean values for the chain and PN-dipole distributions, but the fact that
the distributions are so broad implies a high degree of rotational freedom in the azimuthal angle
for the PN dipole in both phases.

\subsection{2D distributions}

\begin{figure}[h]
\begin{center}
\includegraphics[width=0.85\linewidth,angle=0]{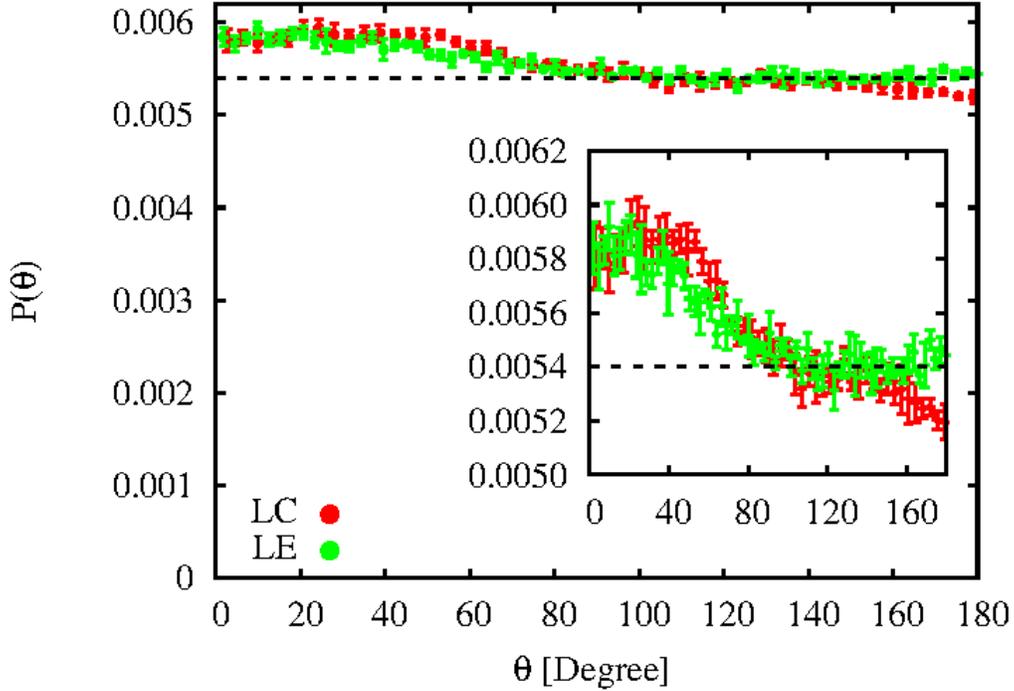}
\caption{\label{fig8} Angle between projected head and tail in LC and LE domains.}
\end{center}
\end{figure}

To confirm the scenario that results from the 3D distributions, we have also calculated
the corresponding distributions of the projected chain and PN-dipole vectors, and of their
relative angle. Even though the projected chains show a high degree of angular order extending
throughout LC domains, see Fig. \ref{fig3}, their distribution is considerably blurred when 
referred to a fixed direction in the monolayer because of the deformations of the vector fields 
within the domains. The PN-dipole distribution suffers from the same problem. 
Therefore we only show the distribution in the relative angle between the projected chain and PN-dipole vectors
in Fig. \ref{fig8} for the LC and LE domains, averaged over the whole coexistence region. Here we 
clearly see that the PN-dipole projection is almost completely disordered
with respect to the well defined local chain orientation, confirming the conclusion drawn from the
3D distributions that PN dipoles exhibit almost full rotational libration in the azimuthal
angle.

\subsection{Correlation functions}

The above results, inferred from angular 2D and 3D distributions,
are confirmed by calculating different correlation
functions. We define $f_k^{(t)}(r)\equiv\left<\cos{k\varphi}\right>(r)$ as a radial correlation 
function, where $\cos{\varphi}=\hat{\bm t}_{\perp}({\bm 0})\cdot\hat{\bm t}_{\perp}({\bm r}_{\perp})$,
$\hat{\bm t}_{\perp}({\bm 0})$ is the projected unit tail vector for a molecule located at 
the origin, and  $\hat{\bm t}_{\perp}({\bm r}_{\perp})$ is the same vector but for a molecule 
at a transverse distance from the latter given by the vector ${\bm r}_{\perp}$.
The angular brackets $\left<\cdots\right>$ mean average over molecules and time.
Therefore, $f_1^{(t)}(r)$ and $f_2^{(t)}(r)$ measure angular correlations in the monolayer. 
Similar functions $f_1^{(m)}(r)$ and $f_2^{(m)}(r)$ can be defined and computed for the projected PN vectors.
Extracting all these four functions separately for the two types of domains is simple, as molecules can
be identified as belonging to one phase or the other. However, due to the finite size of the domains, 
the long- and intermediate-range behaviour of the functions have to be interpreted with care.

\begin{figure}[h]
\begin{center}
\includegraphics[width=1.00\linewidth,angle=0]{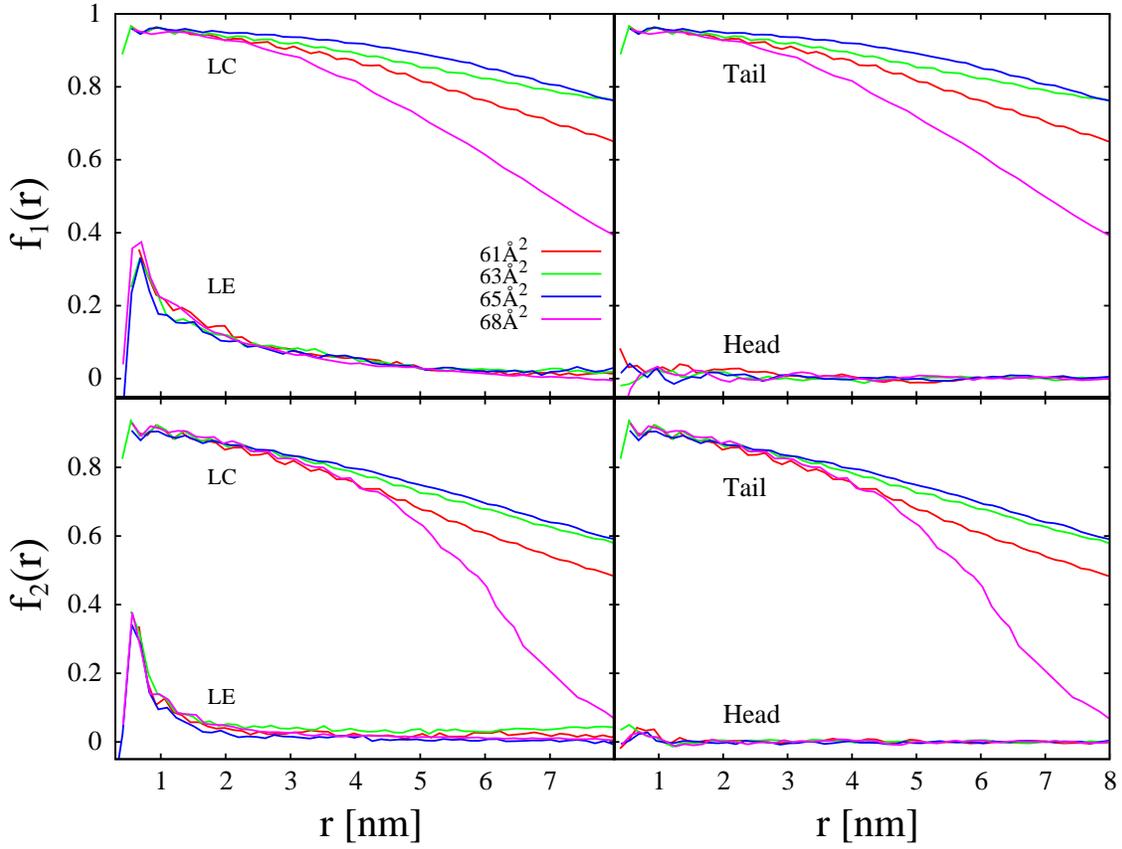}
\caption{\label{Correlation} Correlation functions $f_1(r)$ and $f_2(r)$. (a) and (c): tail
correlation functions $f_1^{(t)}(r)$ and $f_2^{(t)}(r)$ for different
mean areas per lipid and for the two regions LE and LC. (b): $f_1^{(t)}(r)$ and $f_1^{(m)}(r)$
for the LC domains. (d) $f_2^{(t)}(r)$ and $f_2^{(m)}(r)$ for the LC domains. In all cases 
mean areas per lipid are given by the key in (a).}
\end{center}
\end{figure}

Figs. \ref{Correlation} shows these functions for LC and LE regions and for several
values of mean area per lipid. We can see, panels (a) and (c), that long-range correlations are quite evident
in both polar and nematic correlations for the tails in the LC domains,
as expected from the high degree of orientation in LC domains. The functions exhibit
weak oscillations at short distances and stronger quasi-long range order at long distances as 
domains become larger. By contrast, correlations between PN heads rapidly decay with distance, 
panels (b) and (d).
These panels compare the corresponding functions for tails and heads, and show the dramatically
different behaviours. These results are consistent with the conclusions drawn from the angular 
distributions. While one would expect the packed hydrocarbon chains to give rise to strong correlations
at long distances in the LC phase, and water-solvated head groups to exhibit damped behaviour, 
the absence of correlations in the latter is somewhat surprising. 

\subsection{Penetration of lipid molecules into the subphase in LE and LC domains}

\begin{figure}[h]
\begin{center}
\includegraphics[width=0.90\linewidth,angle=0]{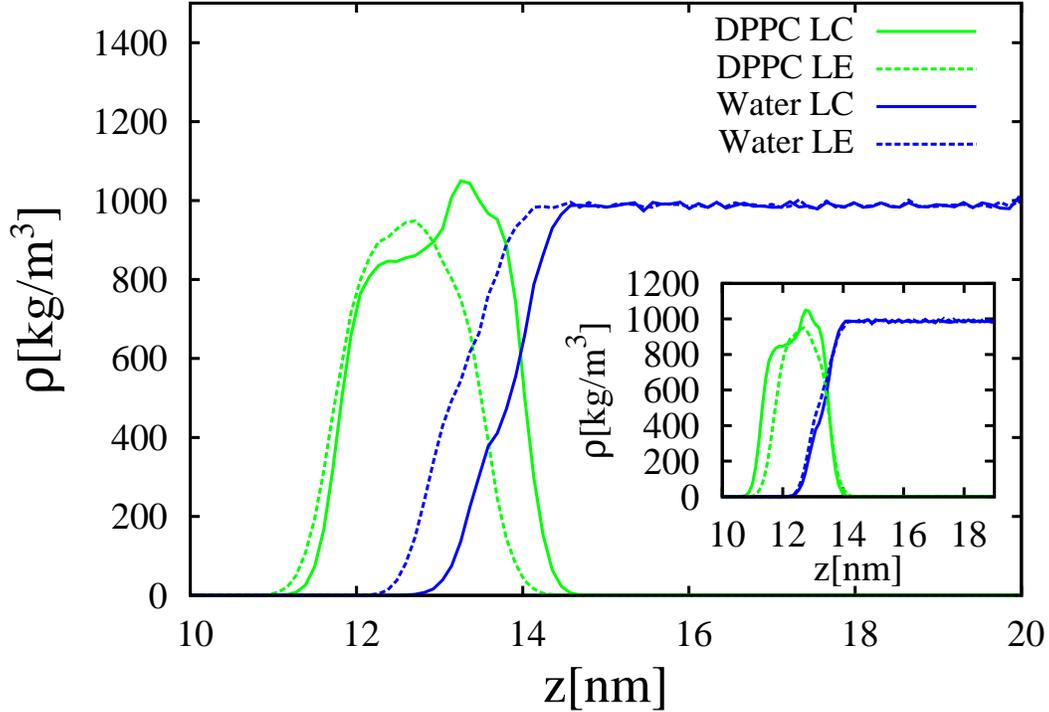}
\caption{\label{Profiles} Mass density profiles of lipid and water molecules corresponding to one of 
the monolayers at a mean area per lipid of $63$ \AA$^2$, for both LC and LE regions. In the inset
the LC profiles have been displaced so as to make the water profiles in both regions coincide.}
\end{center}
\end{figure}

Several authors (see Ref. \cite{Gallegos} and references therein) 
have speculated on the different arrangement of lipid
molecules in the LC and LE phases. In particular, it is believed that the PN dipole is located
deeply into the aquaeous subphase in the LE regions and that, as chains become closely packed
because of the increased lateral pressure, solvation water around dipoles is expelled which
forces the dipoles to become closer to the water surface. In order to investigate this, we have
calculated the density profiles across the interface for the water and lipid molecules
in both phases. These are shown in Fig. \ref{Profiles}, where the densities refer to the whole
mass content of lipid and water molecules separately, including all interaction centres in both
cases. The horizontal axis is the $z$ coordinate in the frame attached to the simulation box.
The profiles correspond to a mean area per lipid of $63$ \AA$^2$, where approximately 50\% of
each monolayer consists of a single LC domain. Density profiles are presently separately for
the two regions, LC and LE. Therefore, the density profile for water corresponds to the water
column below the corresponding domain. Since domains are large, we expect small interface fluctuations
in both LC and LE regions; the fact that the density profiles of water have the same 
shape implies that the rigidity of the interface is similar in both cases.
It can be seen that the lipid distribution in the LC domains develops a shoulder and becomes
wider. This is due to the crystallisation of the molecular chains. But two features of the 
profiles are particularly interesting: (i) the water profiles are very similar in both phases, 
except for a shift in their
position; (ii) the penetration of the molecular polar heads in water is the same in both regions.
The latter feature is more apparent if the LC profiles are rigidly displaced to make both 
water profiles coincide (see inset): the polar heads show no relative change with respect to water.
Our conclusions are that, on the one hand, the polar heads do not approach the water interface
as LC domains are formed at the LE-LC phase transition and, as a consequence, the hydration 
state of the
polar head is not changed. And, on the other hand, the water interface appears to be displaced
by polar heads as they reorganise as a result of molecular chains becoming more closely packed 
in LC domains.

\section{Conclusions}

In this work we have analyzed the coexistence region between LC and LE phases in a DPPC monolayer. 
For the first time, the properties of the coexisting phases have been studied separately by identifying
the molecules that belong to each phase. Moreover, the present simulations involve considerably larger
system sizes than previous works. In practice, simulation times have been 
long enough to allow the system to phase separate completely which ensures that true equilibrium states 
of the monolayer are being analyzed. By contrast, previous works so far have focused on the properties 
of phases without imposing coexistence conditions. This strategy does not avoid the presence of domains 
of the opposite phase which might contaminate the averaged properties of the phase under consideration.

Several conclusions from the present simulation study confirm previous results from other groups.
For example, the angular distribution of the tilt angle of chains is consistent with
the ones reported previously \cite{Dominguez,Choe}. Visual inspection of the projected molecular
shows that there is a high degree of order in LC domains while the distribution is uniform in LE domains.

As for the PN-dipole distributions, we conclude that tilt angle distribution is always unimodal and very 
similar in both phases. The average angle is about $111^{\circ}$, i.e. the PN dipole points $21^{\circ}$
towards the water from the monolayer plane.
This value is much larger than the previously reported value of $\sim 5^{\circ}$, and 
implies a much larger contribution to the molecular dipole perpendicular to the monolayer. Comparison
of this result with previous studies is hindered by the very different thermodynamic states used in
those works. In particular, Mohammad-Aghaie et al. \cite{Bresme} report a bimodal distribution for a 
supposely pure LC phase. Such a bimodal distribution is difficult to understand for a pure phase in 
equilibrium.

In addition, we have also explored the distribution of the angle between the projected chain and 
PN-dipole vectors. To our knowledge this is the first time such an azimuthal distribution is reported.
Assuming a some degree of rigidity in the molecular structure of the lipid molecules, 
we would expect a high correlation
between the orientation of the chains and of the PN head in a monolayer of DPPC molecules.
However, at least within the present force fields, this does not seem to be the case.
According to our results, the polar angle of PN heads has a well defined value, but the head seems to 
be subject to almost
free librations in the azimuthal angle (angle measured in the plane of the monolayer). Further work
on this point may require a more detailed investigation of the reliability of the present force field
and more structural results from the experimental side. The present
results indicate that the component of the dipolar moment parallel to the monolayer, which is
larger than the perpendicular component, would play no role 
since it is completely disordered, even in LC domains. This is in contrast with common wisdom
according to which the PN dipole adopts a close-to-parallel configuration in the LC phase, whereas in
the LE phase it has a more flexible orientation that can change with respect to the packed LC
configuration.

Contrary to earlier suggestions \cite{Gallegos}, our results indicate that lipid 
molecules show the same degree of penetration in the water side of the interface, regardless of the
phase, LE or LC. Instead, strong vdW forces between chains appear to give rise to a displacement
of the whole interface. However, this feature might be a consequence of the force fields used and
the treatment of the head-group hydration, which depends sensitively on force field and
molecular geometry.

LC domains are observed to be elongated,
as discussed by many authors \cite{domains1,domains2,domains3,domains4}. Their shape is not elliptical 
but rather resembles a discorectangle. However longer simulations would be necessary to confirm this issue.
Nonspherical shapes may be induced by competition between
various relevant forces, more importantly vdW forces showing up as a line tension, and forces from
perpendicular dipoles. The strong alignment of lipid chains along domain boundaries may imply the
presence of a significant anisotropic line tension, which should be incorporated to the theoretical
models \cite{McConnell,Aurora}. Also, visual inspection of the domains shows that the sign of the curvature may be an important ingredient in theoretical models.

\acknowledgements

We acknowledge financial support from grants
FIS2017-86007-C3-1-P and FIS2017-86007-C3-2-P from Ministerio de
Econom\'{\i}a, Industria y Competitividad (MINECO) of Spain. 
Work in DPTs groups is supported by the Natural Sciences and Engineering Research Council (Canada), with additional support from the Canada Research Chair Program. Simulations were carried out on Compute Canada resources, funded by the Canada Foundation for Innovation and partners. SP wishes to acknowledge financial support from MINECO through a FPI PhD fellowship. \\
\\
\noindent
S. Panzuela ORCID: 0000-0002-0177-7888\\
\noindent
D. P. Tieleman ORCID: 0000-0001-5507-0688\\
\noindent
L. Mederos ORCID: 0000-0002-4579-8680\\
\noindent
E. Velasco ORCID: 0000-0002-7350-1813


\begin{references}
\bibitem{Perejil1} W\"utnecka, R.; P\'erez-Gil, J.; W\"ustnecka, N.; Cruzbi, A.; Fainermaan V. B.; 
Pison, U. Interfacial Properties of Pulmonary Surfactant Layers. {\it Adv. Colloid Interfasce Sci.}
{\bf 2005}, 117, 33-58.
\bibitem{Perejil2} Parra, E.; P\'erez-Gil, J.
Composition, structure and mechanical properties define performance of pulmonary surfactant 
membranes and films. {\it Chem. Phys. Lipids} {\bf 2008}, 185, 153-175.
\bibitem{Casals} Casals, C; Ca\~nadas, O. 
Role of lipid ordered/disordered phase coexistence in pulmonary surfactant function.
{\it Biochim. Biophys. Acta} {\bf 2012}, 1818, 2550-2562.
\bibitem{Kaganer} Kaganer, V. M.; M\"ohwald, H.; Dutta, P. Structure and Phase Transitions in 
Langmuir Monolayers. {\it Rev. Mod. Phys.} {\bf 1999}, 71, 778-891.
\bibitem{Peter1} Baoukina, S.; Tieleman, D. P. Computer simulations of lung 
surfactant. {\it Biochim. Biophys. Acta-Biomembranes} {\bf 2016}, 1858, 2431-2440.
\bibitem{Peter2} Baoukina, S.; Mendez-Villuendas, E.; Tieleman, D. P. Molecular view of phase
coexistence in lipid monolayers. {\it JACS} {\bf 2012}, 134, 17543-17553.
\bibitem{isotherm} Duncan, S. L.; Larson, R. G. Comparing Experimental and Simulated Pressure-Area 
Isotherms for DPPC. {\it Biophys. J.} {\bf 2008}, 94, 2965-2986.
\bibitem{domains1} Moy, V. T.; Keller, D. J.; Gaub, H. E.; McConnell, H. M. Long-Range Molecular 
Orientational Order in Monolayer Solid Domains of Phospholipid. {\it J. Phys. Chem.} {\bf 1986},
90, 3198-3202.
\bibitem{domains2} Keller, D. J; Korb, J. P.; McConnell, H. M. Theory of Shape Transitions in 
Two-Dimensional Phospholipid Domains. {\it J. Phys. Chem.} {\bf 1987}, 91, 6417-6422.
\bibitem{domains3} McConlogue, C. W.; Vanderlick, T. K. A Close Look at Domain Formation in DPPC 
Monolayers. {\it J. Phys. Chem.} {\bf 1987}, 91, 6417-6422.
\bibitem{domains4} Fl\"orsheimer, M.; M\"ohwald, H. Development of Equilibrium Domain Shapes in 
Phospholipid Monolayers. {\it Chemistry and Physics of Lipids} {\bf 1989}, 49, 231-241.
\bibitem{Mohwald} M\"ohwald, H. Phospolipid and Phospholipid-Protein Monolayers at the Air/Water 
Interface. {\it Annu. Rev. Phys. Chem.} {\bf 1990}, 41, 441-476.
\bibitem{Ma} Ma, G.; Allen, H. C. DPPC Langmuir Monolayer at the Air-Water Interface: Probing the 
Tail and Head Groups by Vibrational Sum Frecuency Generation Spectroscopy. {\it Langmuir} 
{\bf 2006}, 22, 5341-5349.
\bibitem{Rose} Rose, D.; Rendell, J.; Lee, D.; Nag, K.; Booth, V. Molecular Dynamics
Simulations of Lung Surfactant Lipid Monolayers. {\it Biophys. Chem.} {\bf 2008}, 138, 67-77.
\bibitem{Bresme} Mohammad-Aghaie, D.; Mac\'e, E.; Sennoga, C. A.; Seddon, J. M.; Bresme, F.
Molecular Dynamics Simulations of Liquid-Condensed to Liquid-Expanded Transitions in
DPPC Monolayers. {\it J. Phys. Chem. B} {\bf 2010}, 114, 1325-1335.
\bibitem{Huynh} Huynh, L; Perrot, N.; Beswick, V.; Rosilio, V.; Curmi, P. A.; Sanson, A.;
Jamin, N. Structural Properties of POPC Monolayers under Lateral Compression: 
Computer Simulation Analysis. {\it Langmuir} {\bf 2014} 30, 564-573.
\bibitem{Javanainen} Javanainen, M.; Lamberg, A.; Cwiklik, L.; Vattulainen, I.; Samuli Ollila, O. H.
Atomistic Model for Nearly Quantitative Simulations of Langmuir Monolayers.
{\it Langmuir} {\bf 2018}, 34, 2565-2572.
\bibitem{CHARMM} Klauda, J. B.; Venable, R. M.; Freites, J. A.; O'Connor, J. W.; Tobias, D. J.;
Mondragon-Ramirez, C.; Vorobyov, I.; Mackerell, A. D., Jr; Pastor, R. W. Update of the CHARMM
All-Atom Additive Force Field for Lipids: Validation on Six Lipid Types. 
{\it J. Phys. Chem. B.} {\bf 2010}, 114, 7830-7843.
\bibitem{OPC4} Izadi, S.; Anandakrishnan, R.; Onufriev, A. V. Bilding Water Models: A different Approach.
{\it J. Phys. Chem. Lett.} {\bf 2014}, 5, 3863-3871.
\bibitem{GROMACS1} Pronk, S.; P\'all, S.; Schulz, R.; Larsson, P.; Bjelkmar, P.; Apostolov, R.; Shirts, M. R.;
Smith, J. C.; Kasson, P. M.; van der Spoel, D.; Hess, B.; Lindahl, E. GROMACS 4.5: A High-Throughput and 
Highly Parallel Open Source Molecular Simulation Toolkit. {\it Bioinformatics} {\bf 2013}, 29, 845-854.
\bibitem{GROMACS2} Abraham, M. J.; Murtola, T.; Schulz, R.; P\'all, S.; Smith, J. C.; Hess, B.; Lindahl, E. 
GROMACS: High Performance Molecular Simulations Through Multi-Level Parallelism From Laptops to Supercomputers. 
{\it SoftwareX} {\bf 2015}, 1-2, 19-25.
\bibitem{PME} in 't Veld, P. J.; Ismail, A. E.; Grest, G. S. Application of Ewald Summations to Long-Range
Dispersion Forces. {\it J. Chem. Phys.} {\bf 2007}, 127, 144711.
\bibitem{Dominguez} Dominguez, H.; Smondyrev, A. M.; Berkowitz, M.L. Computer Simulations of 
Phosphatidylcholine Monolayers at Air/Water and CCl$_4$/Water Interfaces. {\it J. Phys. Chem. B}
{\bf 1999}, 103, 9582-9588.
\bibitem{Hauser2} Hauser, H.; Phillips, M. C.  in: Danielli, J. F.; Rosenberg, M.D.; 
Cadenhead, D. A.  (Eds.), Progress  in  Surface  and Membrane  Science {\bf 1979}, 13, 297 
(Academic  Press, New York).
\bibitem{Gallegos} Mi\~nones Jr, J.; Rodr\'{\i}guez Patino, J. M.; Conde, O.; Carrera, C.; 
Seoane, R. The Effect of Polar Groups on Structural Characteristics of Phospholipid Monolayers 
Spread at the Air-Water Interface. {\it Col. Surf. Sci. A: Physicochem. Eng. Aspects} {\bf 2002},
203, 273-286.
\bibitem{Sriram} Sriram, I.; Schwartz, D. K. Line tension between coexisting phases in monolayers 
and bilayers of amphiphilic molecules. {\it Surf. Sci. Rep.} {\bf 2012}, 6, 143-159.
\bibitem{Comment} Lamberg, A.; Samuli Ollila, O. H. Comment on `Structural Properties of 
POPC Monolayers under Lateral Compression: Computer Simulation Analysis'. {\it Langmuir} {\bf 2015},
31, 886-887.
\bibitem{Reply} Huynh, L.; Perrot, N.; Beswick, V.; Rosilio, V.; Curmi, P. A.; Sanson, A.;
Jamin, N. Reply to `Comment on "Structural Properties of POPC Monolayers under
Lateral Compression: Computer Simulation Analysis"'. {\it Langmuir} {\bf 2015}, 31, 888-889.
\bibitem{Bresme2} Mohammad-Aghaie, D.; Bresme, F. Force-field Dependence on the 
Liquid-Expanded to Liquid-Condensed Transition in DPPC Monolayers. {\it Mol. Simul.} {\bf 2016},
42, 391-397.
\bibitem{Baoukina} Baoukina, S.; Monticelli, L.; Marrink, S. J.; Tieleman, D. P. Pressure-Area 
Isotherm of a Lipid Monolayer from Molecular Dynamics Simulations. {\it Langmuir} {\bf 2007}, 23,
12617-12623.
\bibitem{Choe} Choe, S.; Chang, R.; Jeon, J.;  Violi, A. Molecular Dynamics Study of a Pulmonary 
Surfactant Film interacting with a Carbonaceous Nanoparticle. {\it Biophys. J.} {\bf 2008}, 95, 
4102-4114.
\bibitem{Hauser} Hauser, H.; Pascher, I.; Pearson, R.; Sundell, S. Preferred Conformation and 
Molecular Packing of Phosphatidylethanolamine and Phosphatidylcholine. 
{\it Biochim. Biophis. Acta} {\bf 1981}, 650, 21-51.
\bibitem{Sun} Sun, F. Constant Normal Pressure, Constant Surface Tension, and Constant
Temperature Molecular Dynamics Simulations of Hydrated 1,2-Dilignoceroylphosphatidylcholine
Monolayer. {\it Biophys. J.} {\bf 2002}, 82, 2511-2519.
\bibitem{McConnell} McConnell, H. M.; Moy, V. T. Shapes of finite two-dimensional lipid domains. 
{\it J. Phys. Chem.} {\bf 1988}, 92, 4520-4525.
\bibitem{Aurora} Campelo, F; Cruz, A.; P\'erez-Gil, J.; V\'azquez, L.; Hern\'andez-Machado, A. 
Phase-field Model for the Morphology of Monolayer Lipid Domains. {\it Eur. Phys. J. E} {\bf 2012},
35, 49.

\end{references}
\end{document}